# TOWARD IMPROVING THE QUALITY OF DOCTORAL EDUCATION: A FOCUS ON STATISTICS, RESEARCH METHODS, AND DISSERTATION SUPERVISION

Rossi A. Hassad
Mercy College, New York, United States of America
Rhassad@mercy.edu

*Doctoral education (PhD) in the USA has long been characterized as being in a crisis, yet empirical research to identify possible determinants is limited, in particular, faculty competence has received only scant research attention. This study ascertained from students, faculty and consultants, their concerns about the teaching of statistics and research (including dissertation supervision). The responses encompass the curriculum, pedagogy, content knowledge, support, and accountability. The current U.S. doctoral education model needs to be systematically reviewed toward assessing its relevance to the changing needs of the disciplines and the job market. In this regard, the almost universal emphasis on evidence-based practice, especially in the disciplines of health and behavioral sciences must be given major consideration. Reform initiatives must also address the roles and qualifications of dissertation committee members (including consultants), the composition of the dissertation committee, and training geared toward preparing and certifying faculty to serve as dissertation committee members.*

INTRODUCTION

Doctoral education (PhD) in the USA is generally characterized as being in a crisis (CGS, 2008; Geiger, 1997; Hamilton, 2003; Kendall, 2002; Meacham, 2002). Evidence abounds of low graduation rates, high attrition rates, protracted dissertation periods, poor quality dissertations, frustrated and unhappy students, and graduates who are inadequately prepared to meet the challenges and needs of their respective disciplines (Golde & Walker, 2006). Specifically, only about 57 percent of students from U.S. institutions complete their PhD programs within 10 years (CGS, 2007). This dismal outcome has been empirically attributed to admission standards, student characteristics, personal and familial circumstances, faculty support, and financial resources (Kendall, 2002; Hamilton, 2003; CGS, 2008). However, only scant research attention has been given to faculty competence, which, in this context, encompasses discipline-specific knowledge, skills and scholarship, as well as commitment to teaching and mentoring in core curricular areas such as statistics and research methods. Perlmutter (2006) in an article titled "Betrayed by Your Adviser" points to the faculty as one of the main "culprits" of underprepared doctoral graduates.

Faculty competence in statistics and research methods is particularly important, as the primary goal of PhD education is to facilitate students to conduct independent and scholarly research (Schreiterer, 2006). In this regard, there is a huge bias toward quantitative dissertation research (Nesselroade & Molenaar, 2003), which is generally viewed as more objective and scientific compared to qualitative research. Enlisting faculty who are competent in the teaching of statistics and research methods is also necessary in order to facilitate evidence-based thinking and practice, particularly in the health and behavioral sciences (including psychology where the doctoral degree is the general entry-level qualification). Evidence-based practice is defined as "*the conscientious, explicit, and judicious use of current best evidence in making decisions about the care of individual patients*" (Sackett et al., 1996, p.71). Such appraisal and use of data necessitate research and statistical competence (Cox, 1997).

OBJECTIVE

The purpose of this study was to ascertain from doctoral students, faculty and consultants, their concerns about the teaching of statistics and research methods (including dissertation supervision), and strategies for improving the quality of doctoral education (PhD programs).

METHODOLOGY

A qualitative research approach was used. Data were collected (in 2009) via online discussion forums (including EDSTAT and ALLSTAT)**,** from faculty members (including PhD dissertation committee members) consultants (research and statistics), doctoral students (and candidates), as well as recent doctoral graduates. Specifically, respondents were asked to share their concerns about the teaching of statistics and research methods (including dissertation supervision), and recommend strategies for improving the quality of doctoral education (PhD programs). There were 25 respondents, and thematic analysis of their open-ended responses was performed. Triangulation of data from multiple groups (students, faculty, recent graduates, and consultants) helped to identify salient issues and concerns.

RESULTS

The following are the core themes from analysis of the qualitative (open-ended) responses along with selected supporting quotes.

1. Greater faculty support is required, in terms of faculty availability and accessibility, as well as possessing relevant expertise, and commitment to doctoral education.

Student: *"I received little support from my supervisor. It seems to me that after discussing with several PhD students, we all shared the same sentiment that our supervisors are too engrossed with their own research and we are not a priority."*

Student: *"I have been very disappointed in the statistical help that my committee provides. My main advisor has very limited information about statistical procedures. He has never heard of data exploration or checking for heteroscedasticity, for example. He just plugs everything into SPSS (or has a research assistant do it), and voila - whatever is spit out, is the final answer."*

Student: *"My committee members are not happy that I have a more complex analysis than they would like to see - just give me the t-test, ANOVA, or correlation, and let's wrap it up, seems to be the prevailing directive. I really learned a lot from the consultant."*

2. Doctoral programs should adopt a more constructivist or integrated curriculum approach (with active and authentic learning strategies), especially with regard to the teaching of statistics and research methods, so that students can experience the core concepts. This approach promotes conceptual understanding, and results in more meaningful learning (and hence transferrable knowledge and skills). Faculty members also expressed concerns about the shrinking of the core curriculum, specifically, with reference to statistics and research methods.

Student: *"I must admit that 90% of what I learned in statistics came from struggling with my data. I found that I forgot basic things until I needed them, asked a stupid question, and then was reminded about what I thought I had learned, but really did not."*

Faculty: *"Our requirements for methodology courses keep getting weaker and weaker and shorter, so students have less and less real TRAINING in how to do things, which clearly shows up in their dissertation research. This is particularly true when committees "allow" students to do projects that are beyond their competency levels."*

3. There seems to be a pervasive lack of faculty support and mentoring in statistics and research design, particularly during the dissertation phase, and although committee members may not possess such expertise, they can be opposed to, or have reservations about students consulting with a statistician and/or methodologist.

Student: *"The other troublesome thing is the view that my advisor has of the "statistics guy". The idea that statisticians don't know how to analyze behavioral data really surprised me. So I think the statistics profession needs more interaction with other departments so they can see the needs, and also be seen as the experts they are."*

Student: *"I have learned that statistics is a specialty. When we are in need of some serious help, we need the consultant. The same for research. It does not signal a failure of professional preparation to do this, especially if the resource is available, and it might further our research."*

Faculty: *"Today there are software, online services, methodology consultants, etc. It is almost impossible to monitor this adequately to ensure that what is presented is THE student's own work."*

Faculty: *"I expect doctoral students to do their own research design, statistics, SPSS analysis, etc. I don't go for the use of statistics consultants, although committee members should play that role in an advisory capacity."*

4. Some faculty members steer students into investigating intricate research questions that require complex research designs and analyses, which can go beyond the curriculum, and expertise of the dissertation committee.

Student: *"I finally used Google, and selected two PhD statisticians. It turned out that my data fit a generalized additive mixed model, and trust me, I had never heard of that before. So the consultant gave me a basic introduction, I found some books on the subject, took an online course, and now we are coming to the end of a fascinating, although expensive journey."*

Faculty: *"The major problem is that the committee pushes the student into asking more complex questions which are natural extensions of the previous work done in a particular field, often times not realizing that this will require more complex analyses as well. This forces students to find people to help them. The result of this is that they often do not understand what was done with the data and are ill prepared to discuss their results. This is a recipe for shoddy dissertations which get approved unless the statistics person (assuming there is one on the committee) objects. Many methodologists would rather just pass it than seem to be the bad guy."*

5. More emphasis needs to be placed on methodological issues, and not just the research findings, in the review of the literature, so that students can better understand, critique, and justify the use of a particular statistical test or research design.

Faculty: *"I impress upon doctoral students that their literature reviews need to examine not just the results of other studies, but also the research methodology and statistical methods that were used, especially given that the purpose of a doctoral program is to produce independent scholars."*

Faculty: *"The literature review MUST not just be content, but process (methodology) as well. If 10 similar studies in your field used chi square, you better have a strong case for doing regression analysis, rather than chi square."*

Faculty: *"I have to take exception because this is a good recipe for never exploring alternative ways to working with data. Far too often, we do see 10 studies using chi square because ...# 1 did it so ... #2 does ... and since #s 1 and 2 did it ... so should #3, and down the line. We get a routine of repetition adopted without necessarily any good rationale for it. Unfortunately, too many journals operate this way too."*

6. The "politics" of education (and the dissertation process) could lead to committee members not thoroughly examining the student during the defense of the proposal and dissertation, in particular, not asking questions relating to statistics and research methods, as these can expose weaknesses, and reflect badly on the committee members.

Faculty: *"In any case, committee members tend to avoid asking questions of students in their proposal meeting or oral defense that would detect if the student really knows or if someone else did the work. Generally speaking, at this stage, we do not want to "fail" the student because it* *looks bad on the committee itself. I am not suggesting that this is how it should be, but just opining that I think this is the way it is."*

CONCLUSION & IMPLICATIONS

This study ascertained information from doctoral students, recent graduates, faculty and consultants, on concerns about faculty competence regarding statistics and research methods (including dissertation supervision), and strategies for improvement. The reported areas of concerns encompass the curriculum, pedagogy, content knowledge, support, and accountability at all stages of the doctoral program. The qualitative design of this study (along with the thematic analysis), the use of selected online discussion forums for data collection, and the number of respondents (n = 25) must be considered when attempting to generalize these results.

The current dominant U.S. PhD education model needs to be systematically reviewed (including curriculum, pedagogy, assessment, faculty role, qualifications, and the composition of the dissertation committee) toward assessing its relevance to the changing needs of the disciplines and the job market. In this regard, the widespread emphasis on evidence-based practice must be given major consideration. Also, it may be an opportune time to explore adopting a PhD education model that is emerging in the European system, and which requires students to produce published peer-reviewed articles instead of (or in addition to) the dissertation. This model adds another layer of accountability (and quality control) to the process, which can facilitate improved faculty support, and result in better prepared graduates. Reform initiatives should also focus on training programs geared toward preparing and certifying faculty to serve as dissertation committee members.